\documentclass[twocolumn,showpacs,preprintnumbers,amsmath,amssymb]{revtex4}
\usepackage{epsfig}
\usepackage{amsmath}

\begin{document}

\title{How Many Nodes are Effectively Accessed in Complex Networks?}

\author{Matheus P. Viana}
\affiliation{Institute of Physics at S\~ao Carlos, University of
S\~ao Paulo, P.O. Box 369, S\~ao Carlos, S\~ao Paulo, 13560-970
Brazil}

\author{Jo\~ao L. B. Batista}
\affiliation{Institute of Physics at S\~ao Carlos, University of
S\~ao Paulo, P.O. Box 369, S\~ao Carlos, S\~ao Paulo, 13560-970
Brazil}

\author{Luciano da F. Costa}
\email{ldfcosta@gmail.com}
\altaffiliation{National Institute of Science and Technology for Complex Systems, Brazil.}
\affiliation{Institute of Physics at S\~ao Carlos, University of
S\~ao Paulo, P.O. Box 369, S\~ao Carlos, S\~ao Paulo, 13560-970
Brazil}

\date{ \today }

\begin{abstract}
The measurement called accessibility has been proposed as a 
means to quantify the efficiency of the communication between nodes 
in complex networks.  This article reports important results regarding 
the properties of the accessibility, including its relationship with the 
average minimal time to visit all nodes reachable after $h$ 
steps along a random walk starting from a source,
as well as the number of nodes that are visited after a finite period
of time.  We characterize the relationship between accessibility and 
the average number of walks required in order to visit all reachable
nodes (the exploration time), conjecture that
the maximum accessibility implies the minimal exploration time, and
confirm the relationship between the accessibility values and the
number of nodes visited after a basic time unit.
The latter relationship is investigated with respect to three
types of dynamics, namely: traditional random walks, self-avoiding
random walks, and preferential random walks.  
\end{abstract}

\maketitle

\section{Introduction}

A critical issue in the study of complex systems regards the
interdependency between connectivity and dynamics 
~\cite{dorogovtsev:2008,barrat:2008,boccaletti:2006}.
For instance, given a specific network topology, it would be interesting
to be able to predict how it would behave with respect to several
types of dynamics.  It has been shown, for example, that
reaction-diffusion dynamics spreads more quickly in scale free complex
networks~\cite{gallos:2004} than in uniformly random networks.  Also, 
consensus dynamics tends to converge faster in small world
topologies~\cite{olfati:2007}.  A possible way to address this problem is to
obtain meaningful measurements of the network topology and then try to
correlate them with relevant properties of the dynamics.  This
analysis can be performed at local or global level, which provide
complementary characterization of the studied relationship between
structure and dynamics.

Particularly important types of dynamics include communications, flow,
and diffusion~\cite{holme:2003,tadic:2007,guimera:2002,kozma:2005}.  
Several real-world complex systems are
underlain by this type of dynamics, including accesses to WWW
pages~\cite{dezso:2006}, disease spreading~\cite{havlin:2010}, power distribution
collapse~\cite{bakke:2006}, underground and highways
systems~\cite{boas:2009,cajueiro:2010}.  Frequently, the activation of these systems
starts at a specific node, or set of nodes -- henceforth called
sources, and unfolds into the remainder of the network in ways that
are intrinsically dependent on the network topology~\cite{ahnert:2009}.  More
specifically, it would be desirable to quantify how effectively a
given source can influence the overall network dynamics.  By
`effectively' it is meant the time that is required for the activation
to reach specific levels at a given set of nodes, or the
total activation at such a set after a given period of time.  These
concepts are closely related to the so-called coupon-collector
problem~\cite{frajolet:1992,boneh:1997}: given a number of coupons
(i.e. nodes), each with a respective probability of occurrence, how
many attempts will be required, in the average, until all coupons are
obtained?  Alternatively, it is also important to identify how many
nodes will be accessed after a given period of time.  The current work
addresses these problems through the concept of
\emph{accessibility}~\cite{access:2008}, which quantifies, for a given source
node, the number of effectively accessible nodes at a given distance
and with respect to a specific dynamics.  In this sense, this measure
complements the traditional hierarchical degree~\cite{silva:2006}, providing
valuable information about the network structure.  Note that the
accessibility takes into account not only the number of nodes at a
given distance, but also the transition probabilities between the
source and these nodes.  

The potential of the accessibility to provide valuable insights about
the structure and dynamics of complex networks has been confirmed 
with respect to many applications (Section ~\ref{app}), including the
definition and identification of the borders of complex
networks~\cite{border}.  However, some important aspects of this
measurement remained to be formalized in a more comprehensive fashion.
For instance, how is the accessibility related to the minimum average
time required for accessing all reachable nodes?  Or, in which sense
does the accessibility quantify the number of effectively accessed
nodes?  To answer these important questions in a satisfying way
constitutes the main objective of the present article, as this paves
the way not only to more complete interpretations of the obtained
results but also to different types of applications and interpretations.
In particular, we show that the accessibility can be interpreted in
conceptually meaningful way as being related to the number of
nodes that can be visited along a given period of time.

This work starts by revising the several applications of the
accessibility already reported in the literature.  Then, we define and
illustrate the accessibility concept, following by establishing the
relationship with the coupon collector problem and showing that the
accessibility is related to the number of nodes effectively accessed
after a period of time.

\section{Applications} 
\label{app}

Several different applications have been reported
by using the accessibility concept. For instance, it has been
shown~\cite{access:2008} that, in geographical networks, nodes located close
to the peripheral regions have lower values of accessibility. By
extending this result to non-geographical networks, it has been possible to
define the border of complex networks as the set of nodes with
accessibility smaller than a given threshold
value~\cite{border}. Moreover, recent investigations have
showed that the position of nodes (inside or outside borders)
drastically affects the activity of nodes~\cite{lucas:2010,viana:2010a}. Other
applications unveiled correlations between the accessibility and
real-world properties of nodes. Particularly, in~\cite{silva:2010} the
authors investigated the network obtained from the theorems in the
Wikipedia. In such a network, each theorem is a node and two nodes are
connected whenever a hyperlink is found between the theorems. The
results indicate that
the older theorems have higher accessibility values, while newer
theorems exhibit lower accessibility values. Consequently, new theorems
are located at the periphery of the network, defining the frontier of
the mathematical knowledge. The accessibility has also been used to
investigate the effects of underground systems on the transportation
properties of large cities. It was showed that overall transportation
can be enhanced by incorporating the underground
networks~\cite{viana:2010}. These results were obtained for the
London and Paris transportation networks.

\section{The effective number of accessible nodes}

Given a source node $i$, suppose it is possible to reach $N_i(h)$
different nodes by performing walks with length $h$ departing from
$i$. Then, we say that $i$ has $N_i$ reachable neighbors at distance 
$h$. Each neighbor is reached with a different probability, which is 
represented by the vector ${\bf
p}_i^{(h)}=\{p_1^{(h)},p_2^{(h)},...,p_{N_i(h)}^{(h)}\}$. Given this
vector, the accessibility of the node $i$, at scale $h$, is defined as:

\begin{equation}
\kappa_i(h) = exp\left({-\sum_jp_j^{(h)}\log p_j^{(h)}}\right).
\label{eq:access}
\end{equation}

\begin{figure}[htb]
  \begin{center}
    \includegraphics{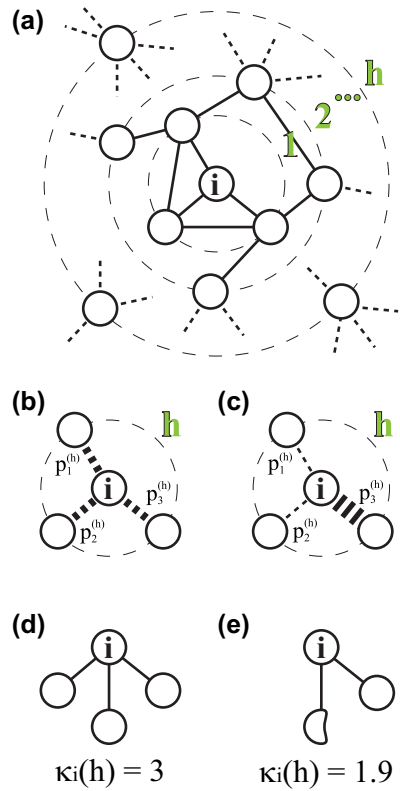}
    \caption{(a) Hierarchical (or concentric) organization around
    the source node (node $i$). (b) Homogeneous case, where all
    neighbors are reached with the same probability. (c) Heterogeneous
    case, where one node has higher probability to be reached. 
    Accessibilities for the (d) homogeneous and (e) heterogeneous
    case.}
	\label{f:sample} 
  \end{center}
\end{figure}

Accessibility values are in the range
$[1,N_i(h)]$, the maximum being obtained for the homogeneous case,
when all probabilities have the same value $1/N_i(h)$. This
measurement, which is related to the heterogeneity of the vector ${\bf
p}$, provides a generalization of the classical concept of
hierarchical (or concentric) degree~\cite{silva:2006},
as explained in Figure \ref{f:sample}. The hierarchical degree of a
source node $i$, at distance $h$ is defined as $k_i(h) = N_i(h)$,
i.e. it is the number of nodes which are at distance $h$ from node $i$. It
is important to note that the value of $k_i(h)$ does not take into
account a dynamical process or respective edge weights in the case of
weighted networks.  The accessibility generalizes the concept of
hierarchical degree by considering that a specific dynamics is unfolding
in the network. We show in this article that the accessibility can be
understood as kind of \emph{effective hierarchical degree}.

In Figure \ref{f:sample}(a), we show the hierarchical levels around
the source node $i$ up to the distance $h$. The network topology, as
well as a type of random walk adopted, will define the transition
probabilities, i.e. the components of the vector $\mathbf{p}$. In
\ref{f:sample}(b) and \ref{f:sample}(c) we represent these
probabilities by using different widths for the edges. Observe that in
both cases, the source node is able to reach $N_i(h)=3$ nodes. In the
first case, all nodes have the same probability, while in the second
case one of the nodes has higher probability than the others. It means
that, in the first case, the source node accesses its
neighbors in a more uniform manner, which yields an accessibility
value equal to 3, as showed in Figure
\ref{f:sample}(d). On the other hand, the interaction between the
source and its neighbors in the second case is biased to a given node,
which decreases the effective hierarchical degree to almost 1.9, as
showed in Figure \ref{f:sample}(e). 

It is important to note that the idea of measuring the heterogeneity 
among first-neighbors nodes in weighted networks was previously proposed 
in ~\cite{almaas:2004,barthelemya:2005}, with the so-called \emph{disparity}. 
More recently, in ~\cite{lee:2010} the authors showed a 
generalization of this measure, namely the \emph{R\'enyi disparity}, which 
is based on the \emph{R\'enyi entropy}. In a particular case, the 
\emph{R\'enyi disparity} uses the \emph{Shannon entropy} in order to quantify 
the heterogeneity of weights attached to the edges of a node . This particular case has 
a similar equation to \ref{eq:access}. However, in our 
case, we consider not only the first-neighbors, but all nodes that can be
reached at distance $h$ by a specific dynamic. In this sense, our approach
can be also applied to non-weighted networks, since we consider the transition
probabilities instead of the edge weights.

Another way to think about the interaction between a source node
and its neighbors is by considering the coupon collector problem. This
problem ~\cite{frajolet:1992,boneh:1997} deals with the following
question: in the average, how many walks with length $h$ departing from
$i$ are required in order to visit all neighboring nodes of $i$ after
$h$ steps at least one time? We will call this quantity
\emph{exploration time} of the node $i$ and denote it by
$\tau_i(h)$, since we can consider the displacement velocity through
the network constant.  Then, the number of walks is proportional to
the time needed to visit all $N_i(h)$ nodes. This problem can be
mapped into a Poisson problem ~\cite{berenbrink:2009} with independent
variables, which yields the expression \ref{eq:ccp_int}.

\begin{equation}
\tau_i(h) = \int_{0}^{\infty}\left(1-\prod_{j=1}^{N_i(h)}\left(1-e^{-xp_j^{(h)}}\right)\right)dx
\label{eq:ccp_int}
\end{equation}

A conjecture has been proposed ~\cite{boneh:1997,caron:1988} that $\tau_i$ reaches 
its minimum value for the homogeneous case, where all neighboring nodes are reached 
with the same probability, i.e. $p_j^{(h)}=1/N_i(h)$ for any $j$. In this case, it is not 
difficult to show that Equation \ref{eq:ccp_int} can be rewritten as:

\begin{equation}
\tau_i^{hom}(h) = N_i(h)\sum_{m=1}^{N_i(h)}\frac{1}{m}.
\end{equation}

\begin{figure}[htb]
  \begin{center}
    \includegraphics{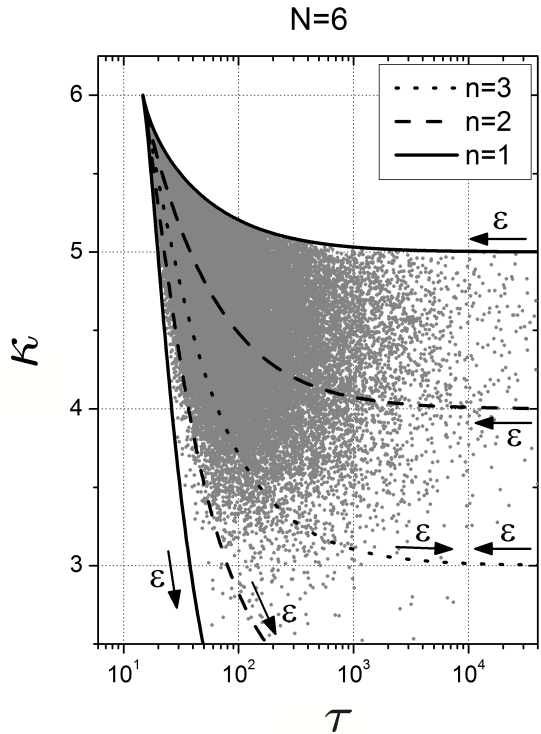}
    \caption{Scatter-plot between the accessibility and the
    exploration time for $10^5$ random vectors with length $N=6$. The
    lines correspond to the cases where the probabilities are divided in two groups
	having the same values among themselves as described by Equations
    \ref{eq:tau_c} and \ref{eq:kappa_c} with parametrization $\epsilon$. }~\label{f:coupon}
  \end{center}
\end{figure}

Therefore, by using the conjecture cited above, we can say that the accessibility 
is maximum whenever the exploration time is minimum. This characteristic is 
illustrated in Figure ~\ref{f:coupon}, which shows a scatter-plot between the 
accessibility, $\kappa$, and the exploration time, $\tau(h)$, for $10^5$ randomly 
generated vectors $\bf{p}$ with length $N=6$. In this plot it is also shown a set of 
important curves, which provides a more comprehensive characterization of the 
probabilities configuration. They correspond to the specific cases where exactly 
$n$ probabilities have a value $\epsilon$, while all the others $(N-n)$ probabilities 
are also identical between themselves (so that the sum of all these probabilities 
becomes equal to one). Therefore, the straight line is related to the case 
where $n=1$, so that $N-1$ probabilities have the same value. Also, this line 
corresponds to the bounding value of the accessibility as a function of $\tau(h)$, 
meaning that all the possible configurations of $\bf{p}$ are enclosed by this curve. 
The dashed line corresponds to the configurations where $n=2$. Similarly, 
the dotted line corresponds to the situations where $n=3$, in this case, 
half of each of the probabilities are equal between themselves. One can use the 
parametrization $\epsilon$ in Equations \ref{eq:access} and \ref{eq:ccp_int} in 
order to obtain a general equation (indexed C) characterizing these curves:

\begin{multline}
	\tau_{C}(\epsilon) = \frac{1}{\epsilon} \sum_{m=1}^{n} \frac{1}{m}  +  \frac{1}{p} \sum_{m=1}^{N-n} \frac{1}{m} - \\
- \sum_{m=1}^{d}\sum_{m'=1}^{N-n} (-1)^{m+m'} {n \choose m}  {{N-n} \choose {m'} }\frac{1}{m\epsilon+m'p}
	\label{eq:tau_c}
\end{multline}

and

\begin{equation}
	\kappa_{C}(\epsilon) = \frac{1}{p} \left(\frac{p}{\epsilon}\right)^{\epsilon n},
	\label{eq:kappa_c}
\end{equation}
where $p=(1-n\epsilon)/(N-n)$. Observe that $\epsilon$ lies in the interval $[0,1/n]$. 
When $\epsilon < 1/N$, the upper part of the curves is obtained. In this case, we have 
$\kappa_C\rightarrow N-n$ and $\tau_C\rightarrow \infty$ for $\epsilon \rightarrow 0$. 
For $\epsilon > 1/N$, we have the bottom part of the curves, for which  
$\kappa_C\rightarrow n$ and $\tau_C\rightarrow \infty$, when $\epsilon \rightarrow 1/n$. 
When $\epsilon=1/N$, we reach the homogeneous case, where the accessibility is 
maximum and the exploration time is minimum.

\subsection{Probabilities in Uniformly Random Networks}

Now we investigate the coupon collector problem in uniformly random networks. More specifically, 
we used 5000 realizations of the Erd\H{o}s-R\'{e}nyi model with 200 nodes and average degree 4, 
and then derived the transition probabilities from these respective networks. We adopted random 
walks originating from each of the nodes in the networks so as to obtain the respective transition 
probabilities (the set of $\bf{p}$'s) by using the powers of the transition matrix \cite{Rieger:2004}.  

\begin{figure}[htb]
  \begin{center}
    \includegraphics[width=0.75\linewidth]{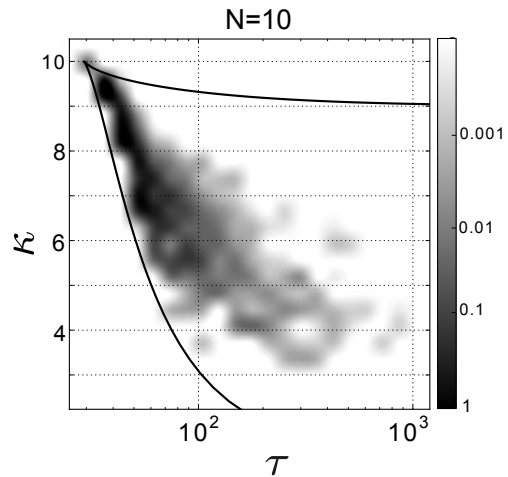}
    \caption{Distribution of the probability configurations in the $\kappa \times \tau$ space 
	obtained for 5000 Erd\H{o}s-R\'{e}nyi networks. We considered
	the cases where 10 nodes are accessible after $h$ steps ($2\leq h \leq15$) 
	along random walks originating from each of the nodes of the networks.}~\label{f:er}
  \end{center}
\end{figure}

Figure \ref{f:er} presents the distribution of the cases in the $\kappa \times \tau$ space. This result 
takes into account all cases where the number of accessible nodes, $N_i(h)$, is equal to 10 for 
values of $h$ in the interval $[2,15]$. The gray levels correspond to the density of cases. Remarkably, 
the density is highly skewed towards the lower bound of the $\epsilon$ curve, and virtually no cases 
are obtained for the upper half of the probabilities region. This means that it 
is extremely unlikely to obtain probability configurations having the majority of nodes with higher 
probability, as illustrated in Figure \ref{f:coupon}.

\begin{figure}[htb]
  \begin{center}
    \includegraphics[width=0.8\linewidth]{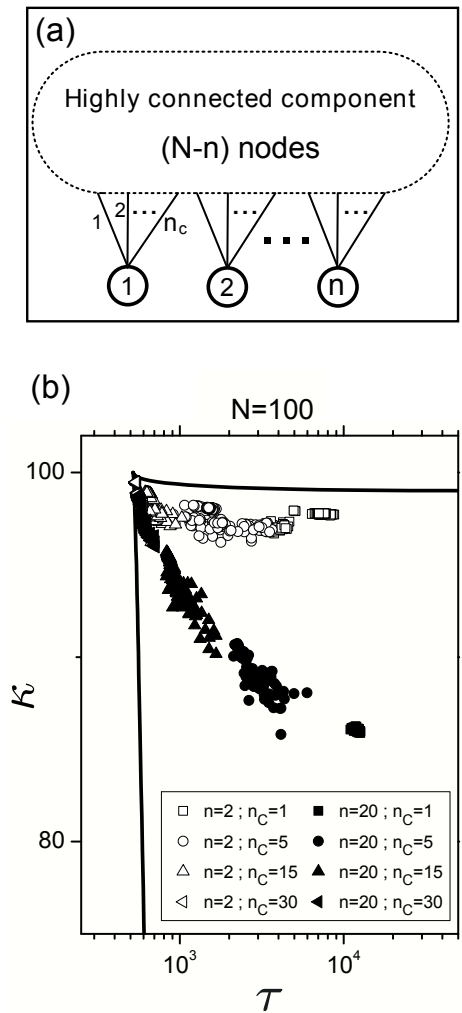}
    \caption{(a) Example of a possible configuration where the upper region of the $\kappa \times \tau$ space
	is occupied. (b) Results obtained for random walks in the considered network for $n=2$ (empty symbols) and
	$n=20$ (filled symbols) and different number of connections $n_c$. }
	\label{f:example} 
  \end{center}
\end{figure}

However, it is possible to obtain configurations which occupy the upper boundary region in the $\kappa \times \tau$ space, where the minority of the probabilities have smaller values. Figure \ref{f:example}(a) presents a particular situation exemplifying this case considering an artificial network with $N$ nodes consisted by two groups: i) a highly connected ER component with $(N-n)$ nodes and average degree $\left< k \right>_c$; and ii) $n$ loosely connected nodes with $n_c$ ($n_c \ll \left< k \right>_c$) links to the previous subgraph. This topological division implies that the nodes in the ER component will be much more accessed than the others when considering random walks in this network, irrespective of the starting node and the length $h$. Thus, in the case of $n \ll N$, the probability vectors, $\mathbf{p}'s$, will have the majority of their components with higher values, thus occupying the upper region of the $\kappa \times \tau$ space. This property is verified for the simulations presented in Figure \ref{f:example}(b) through randomwalks departing from each node for values of $h$ (varied from 2 to 15) where all nodes are reachable. We considered a single realization of the network with $N=100$ and ER component with $\left< k \right>_c$ equal to 50. It was assumed $n=2$ (empty symbols) and $n=20$ (filled symbols) with $n_c$ links, varying from 1 to 30, as indicated in the figure. Observe that, as the value of $n_c$ decreases, the $n$ nodes become less accessible and the points move away from the origin (the homogeneous case), as expected. Although we assumed that the single nodes are directly connected to the ER component, this example can be immediately extended considering the presence of tails of nodes with different sizes. While this network can be artificially created, obtaining similar results for the occupation of the $\kappa \times \tau$ space, it has been showed \cite{costa:2008} that tails are unlikely to occur in great variety of real networks, even for tails with short size. Results for real networks will be shown in the next section.

\begin{figure*}[htb]
  \begin{center}
    \includegraphics[width=0.9\linewidth]{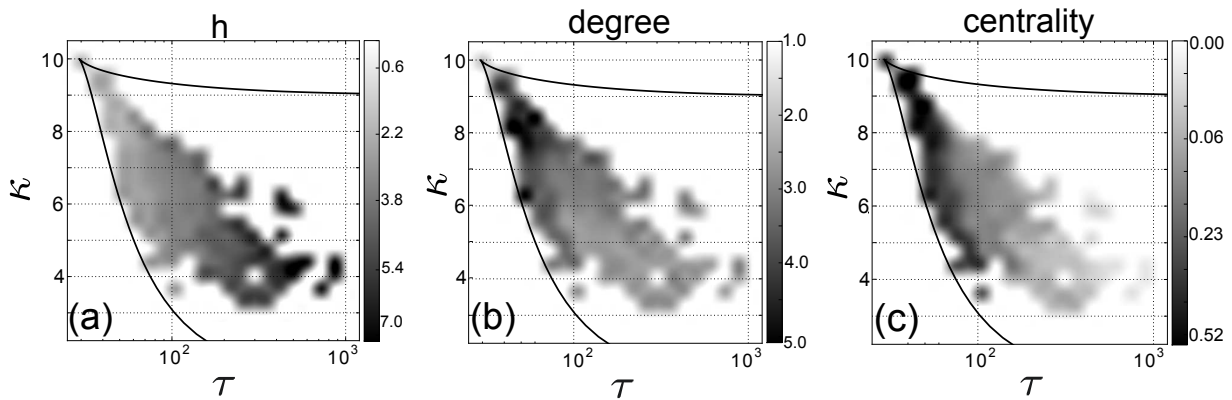}
    \caption{Local average measurements for the probability configurations obtained for 5000 
    Erd\H{o}s-R\'{e}nyi networks. (a) number of steps necessary to reach 10 nodes after departing 
    from the source node, (b) the degree of the source node, and (c) the eigenvector centrality of 
    the source node.}~\label{f:measurements}
  \end{center}
\end{figure*}

Figure \ref{f:measurements} complements the characterization of the $\kappa \times \tau$ space. 
It shows: (a) the local average number of steps necessary to reach 10 nodes after departing from the 
source node; (b) the degree of the source node; and (c) its eigenvector centrality obtained for the 
probability configurations. It is clear from Figure \ref{f:measurements}(a) that random walks with 
larger number of steps (i.e. $h$) tend to have smaller accessibility and longer exploration time. 
On the other hand, random walks starting from nodes with larger degree (Figure \ref{f:measurements}(b)) 
tend to have larger accessibility and shorter exploration times, though in a less definite fashion than 
that observed in Figure \ref{f:measurements}(a). Furthermore, Figure \ref{f:measurements}(c) shows 
a remarkable centrality pattern: it tends to increase with $\kappa$ while decreasing 
with $\tau$, apparently following the level set curves in Figure \ref{f:coupon}.  It should be observed
that these results are specific for the uniformly random ER networks, in the sense that
different trends may be obtained for other theoretical network models.

\subsection{Real-World Probabilities}

\begin{figure}[t!]
  \begin{center}
    \includegraphics[width=0.8\linewidth]{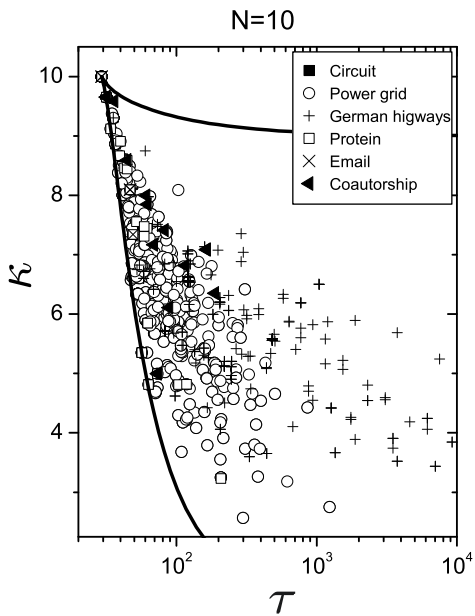}
    \caption{Scatter-plot between the accessibility and the exploration time for 6 real networks 
    considering the cases where $N=10$ nodes are reached after $h = \{ 2, 3, \ldots, 20\}$. }  \label{f:real}
  \end{center}
\end{figure}

We also considered probabilities obtained from real-world networks, namely circuits \cite{millo:2004}, 
power grids \cite{watts:1998}, German highways \cite{kaiser:2004}, protein interactions 
\cite{palla:2005}, e-mails \cite{guimera:2003}, and co-authorships in network science \cite{newman:2006}. 
The probability configurations obtained from these networks are shown in Figure \ref{f:real}. Again, 
most of the cases tend to appear near the lower boundary in the $\kappa \times \tau$ space, which 
is characterized by low exploration time and varying accessibility. This is particularly surprising, as 
it suggests a universal asymmetry in real networks in which a few probabilities are larger than the
others in most configurations. Therefore, it is interesting to observe that real networks, as with the 
uniformly random topologies, tend to minimize the exploration time at the expense of varying 
accessibilities.

\begin{figure*}[htb]
  \begin{center}
    \includegraphics[width=0.9\linewidth]{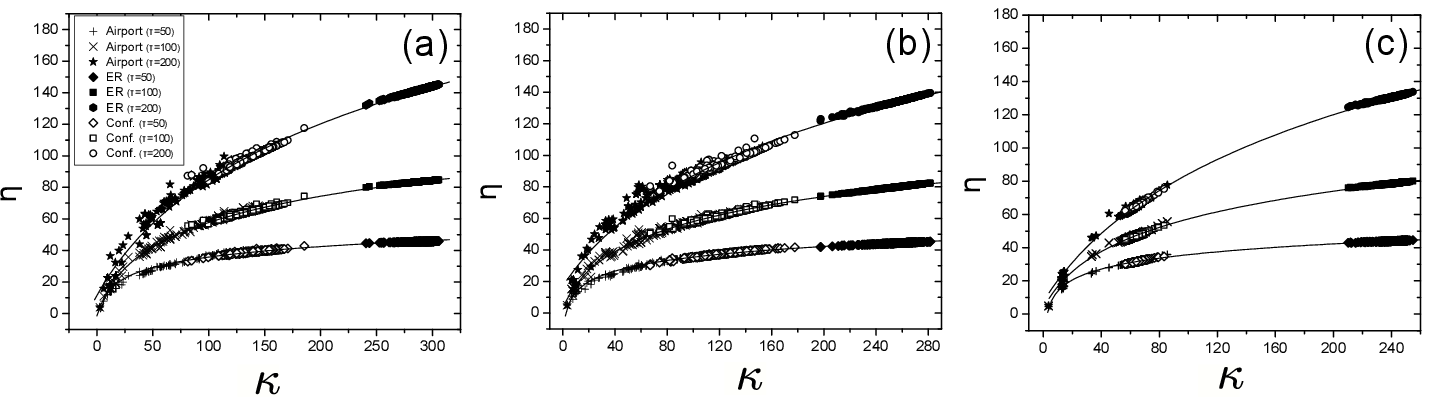}
    \caption{Fraction of reached nodes $\eta_i(t,h)$ as a function of the node accessibility
    $\kappa_i$ for (a) classical random walk, (b) self-avoid random walk and (c) preferential random 
    walk. The results are illustrated for the US airlines network and two random counterparts: 
    \emph{Erd\H{o}s-R\'{e}nyi model} and the respective \emph{Configuration model}~\cite{molley:1995}. All of 
    them have the same number of nodes ($N=332$) and the same average degree 
    ($\langle k \rangle = 12.81$).}  \label{f:relation}
  \end{center}
\end{figure*}

Now, we proceed to a related problem in which we are interested to
know how many nodes, in the average, are visited during the time interval
$t$, while performing a specific type of random walk. This quantity
will be denoted by $\eta_i(t,h))$ and it provides information about
how the network topology around the source node affects the
interaction with its neighbors. After a long time, we expect that
the source node will be able to visit  all $N_i(h)$
neighbors, i.e. $\lim_{t\rightarrow \infty} \eta_i(t,h))\rightarrow
N_i(h)$, independent of the vector ${\bf p}_i^{(h)}$. Therefore, we
can consider that the value of $\eta_i(t,h)$ provides an estimate of
the average number of visited nodes during a finite time. This is
confirmed in Figure ~\ref{f:relation} for the US airlines network
\cite{pajek} and two random counterparts:
\emph{Erd\H{o}s-R\'{e}nyi model} and \emph{Configuration
model}~\cite{molley:1995}.  In order to obtain the transition
probabilities, we considered three different types of random walks:
(i) traditional random walk (TRW), (ii) preferential random walk (PRW)
and (iii) self-avoid random walk (SARW). They were estimated for
$h=\ell$, where $\ell=3$ is the network diameter. The TRW and PRW
dynamics were calculated by using powers of the transition matrix, while
the SARW was estimated through agent-based simulation. This calculation
was repeated $10^6$ times for each source node. It should be noted
that, in the case of SARW dynamics, if an agent cannot proceed
further, it remains at the final node contributing to the
probabilities for the next steps ~\cite{access:2008}. Then, the
transition probabilities were used in the equation ~\ref{eq:access} to
evaluate the accessibility of the node $i$. The values of
$\eta_i(t,h)$ were obtained for each node $i$ as follows: first we
draw $t$ neighbors of node $i$ at distance $h=\ell$ according to the
obtained probabilities $p_j(\ell)$. Then we count how many different
nodes were drawn. The average over several realizations gives us an
estimate of $\eta_i(t,h)$. The behavior of $\kappa$ versus $\eta$ is
showed in Figure \ref{f:relation}a, \ref{f:relation}b and
\ref{f:relation}c for TRW, SARW and PRW respectively. The results have 
been found to be well-fitted by the functional form represented in
Equation \ref{eq:fit}.

\begin{equation}
	\eta = a + b\log(\kappa+c) \label{eq:fit}
\end{equation}

It is interesting to observe in Figure \ref{f:relation}c that when the
preferential rules were adopted, the fraction of reached nodes was
strongly decreased in the US airline and its configuration model. This
is a direct consequence of the degree heterogeneity of these
networks. For preferential random walk, the walks are biased to pass
through nodes with high connectivity, so that several possible paths
are not used. Therefore, the effective number of reached nodes is
smaller than in the cases where we considered self-avoid or
traditional random walk. As we can see, this phenomena is not observed
in ER networks, which have a more homogeneous degree distribution.

\section{Conclusions}

The accessibility concept was introduced recently \cite{access:2008} as a means to quantify 
the potential of a node to interact with other nodes in a complex network. 
Given the many promising results obtained so far, it became important to 
better understand the accessibility concept, specially regarding optimization 
aspects. The present work focused on the investigation of the accessibility 
regarding the coupon collector problem as well as its relationship with the 
average number of nodes visited along a random walk during a given time interval.  

A number of remarkable results were obtained about the relationship 
between the accessibility and the exploration time. First, we have that the minimal 
exploration time is obtained for the maximum accessibility. No relationship between 
these two properties have been observed otherwise, i.e. when we consider all possible 
probabilities configurations. However, in the cases of uniformly random and 
real-world networks, a stronger correlation is verified, with the cases tending to 
lie near the lower boundary in the $\kappa \times \tau$ space. As a matter of fact, 
there is a very low probability of having cases occupying the upper half portion of this space.  
Although this could suggest some intrinsic impossibility of having such cases, we 
showed at least one type of topology leading to a configuration lying over the upper boundary. 
This remarkable result shows that in all considered networks the transition probability
configurations tend to be characterized by small exploration time at the expense of varying accessibilities. 

Regarding the relationship between the accessibility and the average number of nodes 
visited along a random walk during a given time interval, we  showed that the concept of 
accessibility can be understood as a generalization of the classical degree, in the sense 
that the accessibility quantifies the effective number of nodes that can be reached from 
the source node after a given number of steps. In order to confirm this statement, we also 
showed that there is a strong relationship between the accessibility and the inverse 
coupon collector problem, which deals with the number of visited nodes in a finite time interval. 

Future works could take into a account activations originating from multiple nodes as 
well as how other dynamical properties can be predicted from the accessibility values. 
It would be particularly interesting to identify more general theoretical models and real
networks capable of covering the $\kappa \times \tau$ more uniformly.

\begin{acknowledgments}
Luciano da F. Costa is grateful to FAPESP (05/00587-5) and CNPq (301303/06-1 and 
573583/2008-0) for the financial support. M. P. Viana is grateful to FAPESP sponsorship 
(proc. 07/50882-9). J. L. B. Batista thanks CNPq (131309/2009-9) for sponsorship.  
The authors are also grateful to Yoshiharu Kohayakawa (IME-USP) for help regarding 
the coupon-collection problem.
\end{acknowledgments}

\bibliography{viana11access}

\begin{thebibliography}{10}

\bibitem{dorogovtsev:2008}
S.~N. Dorogovtsev and A.~V. Goltsev.
\newblock Critical phenomena in complex networks.
\newblock {\em Rev. Mod. Phys.}, 80(4):1275--1335, 2008.

\bibitem{barrat:2008}
A.~Barrat, M.~Barthélemy, and A.~Vespignani.
\newblock {\em Dynamical Processes on Complex Networks}.
\newblock Cambridge University Press, 2008.

\bibitem{boccaletti:2006}
S.~Boccaletti, V.~Latora, Y.~Moreno, M.~Chavez, and D.~U. Hwang.
\newblock Complex networks: structure and dynamics.
\newblock {\em Physics Reports}, 424(4-5):175--308, 2006.

\bibitem{gallos:2004}
L.~K. Gallos and P.~Argyrakis.
\newblock Absence of kinetic effects in reaction-diffusion processes in
  scale-free networks.
\newblock {\em Phys. Rev. Lett.}, 92(13):138301, 2004.

\bibitem{olfati:2007}
R.~Olfati-Saber, J.~A. Fax, and R.~M. Murray.
\newblock Consensus and cooperation in networked multi-agent systems.
\newblock {\em Proceedings of the IEEE}, 95(1):215--233, 2007.

\bibitem{holme:2003}
P.~Holme.
\newblock Congestion and centrality in traffic flow on complex networks.
\newblock {\em Advances in Complex Systems}, 6(2):163--176, 2003.

\bibitem{tadic:2007}
B.~Tadic, G.J. Rodgers, and S.~Thurner.
\newblock Transport on complex networks: Flow, jamming and optimization.
\newblock {\em Int. J. Bifurcation and Chaos}, 17(7):2363--2385, 2007.

\bibitem{guimera:2002}
R.~Guimera, A.~Arenas, A.~Diaz-Guilera, and F.~Giralt.
\newblock Dynamical properties of model communication networks.
\newblock {\em Phys. Rev. E}, 66(2), 2002.

\bibitem{kozma:2005}
B.~Kozma B, M.~B. Hastings, and G.~Korniss.
\newblock Diffusion processes on power-law small-world networks.
\newblock {\em Phys. Rev. Lett}, 95(1):018701, 2005.

\bibitem{dezso:2006}
Z.~Dezso, E.~Almaas, A.~Lukacs, B.~Racz, I.~Szakadat, and A.~L. Barabasi.
\newblock Dynamics of information access on the web.
\newblock {\em Phys. Rev. E}, 73(6), 2006.

\bibitem{havlin:2010}
M.~Kitsak, L.~K. Gallos, S.~Havlin, F.~Liljeros, L.~Muchnik, H.~E. Stanley, and
  H.~A. Makse.
\newblock Identification of influential spreaders in complex networks.
\newblock {\em Nature Physics}, 6(11):888--893, 2010.

\bibitem{bakke:2006}
J.~O.~H. Bakke, A.~Hansen, and J.~Kertesz.
\newblock Failures and avalanches in complex networks.
\newblock {\em Europhys. Lett.}, 76(4):717--723, 2006.

\bibitem{boas:2009}
P.~R.~V. Boas, F.~A. Rodrigues, and L.~da~F.~Costa.
\newblock Modeling worldwide highway networks.
\newblock {\em Physics Letters A}, 374(1):22--27, 2009.

\bibitem{cajueiro:2010}
D.~O. Cajueiro.
\newblock Optimal navigation for characterizing the role of the nodes in
  complex networks.
\newblock {\em Physica A: Statistical Mechanics and its Applications},
  389(9):1945--1954, 2010.

\bibitem{ahnert:2009}
S.~E. Ahnert, B.~A.~N. Traven\c{c}olo, and L.~da~F.~Costa.
\newblock Connectivity and dynamics of neuronal networks as defined by the
  shape of individual neurons.
\newblock {\em New J. Phys.}, 11:103053, 2009.

\bibitem{frajolet:1992}
P.~Frajolet, D.~Gardi, and L.~Thimonier.
\newblock Birthday paradox, coupon collectors, caching algorithms and
  self-organizing search.
\newblock {\em Discrete Applied Mathematics}, 39:207--229, 1992.

\bibitem{boneh:1997}
A.~Boneh and M~Hofri.
\newblock The coupon-collector problem revisited - a survey of engineering
  problems and computational methods.
\newblock {\em Stochastic Models}, 13(1):39--66, 1997.

\bibitem{access:2008}
B.~A.~N. Travençolo and L.~da~F.~Costa.
\newblock Accessibility in complex networks.
\newblock {\em Phys. Lett. A}, 373(1):89--95, 2008.

\bibitem{silva:2006}
F.~N. Silva and L.~da~F.~Costa.
\newblock Hierarchical characterization of complex networks.
\newblock {\em Journal of Statistical Physics}, 125(4):841--872, 2006.

\bibitem{border}
B.~A.~N. Travençolo, M.~P. Viana, and L.~da~F.~Costa.
\newblock Border detection in complex networks.
\newblock {\em New J. Phys.}, 11:063019, 2010.

\bibitem{lucas:2010}
L.~Antiqueira and L.~da~F.~Costa.
\newblock Structure-dynamics interplay in directed complex networks with border
  effects. in: Complenet 2010.
\newblock {\em Proceedings of the 2nd Workshop on Complex Networks.}, 2010.

\bibitem{viana:2010a}
M.~P. Viana, B.~A.~N. Travençolo, E.~Tanck, and L.~da~F.~Costa.
\newblock Characterizing topological and dynamical properties of complex
  networks without border effects.
\newblock {\em Physica A: Statistical Mechanics and its Applications},
  389(8):1771--1778, 2010.

\bibitem{silva:2010}
F.~N. Silva, B.~A.~N. Travençolo, M.~P. Viana, and L.~da~F.~Costa.
\newblock Identifying the borders of mathematical knowledge.
\newblock {\em J. Phys. A: Math. Theor}, 43(32), 2010.

\bibitem{viana:2010}
L.~da~F.~Costa, B.~A.~N. Travençolo, M.~P. Viana, and E.~Strano.
\newblock On the efficiency of transportation systems in large cities.
\newblock {\em Europhys. Lett.}, 91(1), 2010.

\bibitem{almaas:2004}
E.~Almaas, B.~Kovács, T.~Vicsek, Z.~N. Oltvai, and A.-L. Barabási.
\newblock Global organization of metabolic fluxes in the bacterium escherichia
  coli.
\newblock {\em Nature}, 427:839--843, 2004.

\bibitem{barthelemya:2005}
M.~Barthélemya, A.~Barratb, R.~Pastor-Satorras, and A.~Vespignani.
\newblock Characterization and modeling of weighted networks.
\newblock {\em Physica A: Statistical Mechanics and its Applications},
  346(1-2):34--43, 2005.

\bibitem{lee:2010}
S.~H. Lee, P.~J. Kim, Y.~Y. Ahn, and H.~Jeong.
\newblock Googling social interactions: Web search engine based social network
  construction.
\newblock {\em PLOS ONE}, 5(7):e11233, 2010.

\bibitem{berenbrink:2009}
P.~Berenbrink and T.~Sauerwald.
\newblock The weighted coupon collector's problem and applications.
\newblock In {\em Proceedings of the 15th Annual International Conference on
  Computing and Combinatorics}, COCOON '09, pages 449--458. Springer-Verlag,
  2009.

\bibitem{caron:1988}
R.~J. Caron, M.~Hlynka, and J.~F. McDonald.
\newblock On the best-case performance of probabilistic methods for detecting
  necessary constraints.
\newblock {\em Technical Report WMSR-88-02}, 1988.

\bibitem{Rieger:2004}
J.~D. Noh and H.~Rieger.
\newblock Random walks on complex networks.
\newblock {\em Phys. Rev. Letts.}, 92:118701, 2004.

\bibitem{costa:2008}
P.~R. Villas-Boas, F.~A. Rodrigues, G.~Travieso, and L.~da~F.~Costa.
\newblock Chain motifs: The tails and handles of complex networks.
\newblock {\em Phys. Rev. E}, 77(2):026106, 2008.

\bibitem{millo:2004}
R.~Milo, S.~Itzkovitz, N.~Kashtan, R.~Levitt, S.~Shen-Orr, I.~Ayzenshtat,
  M.~Sheffer, and U.~Alon.
\newblock Superfamilies of evolved and designed networks.
\newblock {\em Science}, 303:1538--1542, 2004.

\bibitem{watts:1998}
D.~J. Watts and S.~H. Strogatz.
\newblock Collective dynamics of 'small-world' networks.
\newblock {\em Nature}, 393:440--442, 1998.

\bibitem{kaiser:2004}
M.~Kaiser and C.~C. Hilgetag.
\newblock Spatial growth of real-world networks.
\newblock {\em Phys. Rev. E}, 69:036103, 2004.

\bibitem{palla:2005}
G.~Palla, I.~Derényi, I.~Farkas, and T.~Vicsek.
\newblock Uncovering the overlapping community structure of complex networks in
  nature and society.
\newblock {\em Nature}, 435:814--818, 2005.

\bibitem{guimera:2003}
R.~Guimera, L.~Danon, A.~Diaz-Guilera, F.~Giralt, and A.~Arenas.
\newblock Self-similar community structure in a network of human interactions.
\newblock {\em Phys. Rev. E}, 68:065103, 2003.

\bibitem{newman:2006}
M.~E.~J. Newman.
\newblock Finding community structure in networks using the eigenvectors of
  matrices.
\newblock {\em Phys. Rev. E}, 74:036104, 2006.

\bibitem{molley:1995}
M.~Molloy and B.~Reed.
\newblock A critical point for random graphs with a given degree sequence.
\newblock {\em Random Struct. Algorithms}, 6:161--179, 1995.

\bibitem{pajek}
B.~Vladimir and M.~Andrej.
\newblock Pajek datasets., 2006.

\end{thebibliography}

\end{document}